\documentclass[aps,prl,preprint,groupedaddress]{revtex4}

\usepackage{graphicx}
\begin{document}


\title{Photoemission Electron Microscopy as a tool for the investigation of optical near fields}


\author{M. Cinchetti, A. Gloskovskii, S. A. Nepjiko
\ and G. Sch\"{o}nhense }
\affiliation{Johannes Gutenberg-Universit\"{a}t, Institut f\"{u}r  Physik, 55099 Mainz, Germany}

\author{H. Rochholz}
\author{M. Kreiter}
\email{kreiter@mpip-mainz.mpg.de} \affiliation{Max Planck Institut f\"ur Polymerforschung, 55128 Mainz,
 Germany}


\date{\today}

\begin{abstract}
Photoemission electron microscopy was used to image the electrons photoemitted from specially tailored Ag nanoparticles deposited on a Si
substrate (with its native oxide SiO$_{x}$). Photoemission was induced by illumination with a Hg UV-lamp (photon energy cutoff
$\hbar\omega_{UV}=5.0$\,eV, wavelength $\lambda_{UV}=250$\,nm) and with a Ti:Sapphire femtosecond laser ($\hbar\omega_{l}=3.1$\,eV,
$\lambda_{l}=400$\,nm, pulse width below  200\,fs), respectively. While homogeneous photoelectron emission from the metal is observed upon
illumination at energies above the silver plasmon frequency, at lower photon energies the emission is localized at tips of the structure. This
is interpreted as a signature of the local electrical field therefore providing a tool to map the optical near field with the resolution of
emission electron microscopy.
\end{abstract}


\maketitle

The intensity of optical fields may be largely enhanced in the vicinity of nanoscopically structured metal objects. Extreme local field
enhancement is believed to be responsible for the increase of the Raman cross section of organic molecules by up to a factor of $10^{14}$
\cite{Nie97} in the vicinity of stochastically roughened silver films. Fluorescence as well is drastically altered by an enhanced optical near
field  which was shown to improve the performance of chromophores \cite{Wokaun85, Lakowicz04} and semiconductor quantum dots \cite{Shimizu02}
significantly. These effects are in general explained in terms of an increased coupling of a local absorbing or emitting dipole with both
incident and outgoing far field photons, such affecting both the optical excitation and emission process. In analogy to antenna used for
radiation of lower frequency, in this context metal objects can be regarded as antenna for the optical regime. These nanoscopic antenna are
characterised by an overall optical resonance similar to the well-known plasmon resonance of spherical metal particles \cite{Bohren83,
Kreibig92}. Such resonances have been observed for rods \cite{Soennichsen02}, closely spaced particle dimers \cite{Okamoto03} and nanorings
\cite{Aizpurua03}, to name only a few. In all these cases a strong dependence of the resonance wavelength on the geometry was found. An
important additional requirement for a good antenna are geometrical features of very small dimension that focus the optical field to extremely
high intensities in volumes far below the diffraction limit. One prominent example is the nanoscopic gap which is formed between two almost
touching metal spheres \cite{Aravind81} or cylinders \cite{Kottmann01DoppelZylinder} or between a plane and a sphere \cite{Aravind1982}. Sharp
tips are another important example for nanoscale structures where it could be shown experimentally \cite{Kramer02} that the photophysics of a
single fluorescent molecule is significantly altered by the large field enhancements, this result being in qualitative agreement with theory
\cite{Kottmann01PRB}.

Mapping of the near field distribution down to the nanometer scale is the key to understand and optimize such antenna structures. Fluorescence
microscopy \cite{Ditlbacher2002} has been used for the mapping of near fields but this method is restricted to the structure of the near field
above the diffraction limit. Near field optical microscopy has proven to give a resolution down to some 10 nm \cite{Frey02}. It must be pointed
out, though, that metal tips which provide a good optical resolution form, if approaching the object under study, a highly complex metal
structure which is composed of tip and sample, possessing a nanoscopic gap between tip and object. The optical response is in turn strongly
altered preventing the optical characterization of the sample object alone. Dielectric tips are less perturbing but have only limited
resolution. They have been used to investigate the optical near fields of several plasmonic nanostructures \cite{Krenn1999}. The use of almost
pointlike probes for the near field such as a single molecule \cite{Michaelis00} or the end of a carbon nanotube \cite{Hillenbrand2003}
circumvents the aforementioned problems but the experimental difficulties prevent these approaches from being applicable as a standard method
for the investigation of a larger amount of samples. An approach to use the high resolution of electron microscopy was demonstrated by Yamamoto
et al. \cite{Yamamoto01} who detected light emission induced by an electron beam and were able to image multipolar patterns on small silver
particles, taking advantage of the superior resolution of electron optics.

Photoelectron emission has been shown to be enhanced by the increase of the local electrical field upon excitation of the particle plasmon of
small silver clusters \cite{Lehmann2000}. In this context, photoemission dynamics of surface-bound silver particles was studied in detail
\cite{Pfeiffer2004} and the influence of the collective electron dynamics could be quantified \cite{Scharte2001, Merschdorf2004} as well as
charging of the particles \cite{Pfeiffer2004}. However, in the above mentioned works the photoemission signal was recorded without lateral
resolution. In the present paper it is shown that photoemission electron microscopy (PEEM) can be used to map the near field distribution. Using
the same experimental technique, we have been previously able to localize regions on inhomogeneous Ag and Cu surfaces where laser illumination
excites collective electron modes, or localized surface plasmons (LSP). Laterally resolved electron energy distribution spectra have shown that
the LSP-induced enhanced near field affects the photoemission and its dynamics in a crucial way \cite{Cinchetti04,Cinchetti04b}. In these
experiments the structural features giving rise to the locally enhanced photoemission yield were not known and remained speculative. For this
reason, we present in this publication experiments on well-defined metal structures that posses sharp tips as required for high local field
enhancement.

\begin{figure}
\begin{center}
\includegraphics[width=3.6cm,keepaspectratio=true]{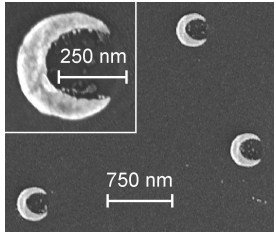}
\caption{Scanning electron micrograph of the silver crescents. The inset shows a magnified view} \label{SEM}
\end{center}
\end{figure}

Nanoscopic crescent-shaped silver objects were prepared on a Si wafer  using a combination of colloid templating, metal film deposition and ion
beam milling \cite{Shumaker2005}. Figure \ref{SEM} shows a typical Scanning Electron Microscopy (SEM) image of the crescents with a diameter of
roughly $400~\mathrm{nm}$ and a thickness of $50~\mathrm{nm}$.
Photoelectron emission was induced by illuminating the sample with two different light sources.
 A Hg deep-UV lamp (photon energy cutoff $\hbar\omega_{UV}=5.0$\,eV,
 wavelength $\lambda_{UV}\geq 250$\,nm), was focused on the sample at an angle
$\theta=65^{\circ}$ with respect to the surface normal. The fundamental of a femtosecond Ti:Sapphire laser (MaiTai Spectra Physics, wavelength
tunable between 750\,nm and 850\,nm, repetition rate 80\,MHz) was frequency-doubled by a commercial device (3980 Spectra Physics), giving a
photon energy tunable between $2.9$ and $3.3$\,eV and a pulse width below $200$\,fs. For this experiment the photon energy was kept fixed at
$\hbar\omega_{l}=3.1$\,eV ($\lambda_{l}=400$\,nm). The frequency-doubled beam was focused on the sample at $\theta=65^{\circ}$, from the
opposite direction in the same plane of incidence as the UV lamp. The obtained fluence per pulse was about $6.4$\,$\mu$J\,cm$^{-2}$. A Fresnel
rhomb allowed to adjust the direction of the polarization vector ($p$-polarization). The photoemission electron microscope was a commercial
instrument (Focus IS-PEEM). Model calculations were performed in order to illustrate the dependence of the optical response of silver on photon
energy. For the calculations, the optical response of silver was described by literature values \cite{Johnson72} and a two-dimensional geometry
of an infinitely extending rod with a cross section similar to the crescents was considered. Maxwell equations were solved with a commercial
Finite Element Code (Femlab GMBH, G\"ottingen).

In a PEEM image, the brightness in a given area is proportional to the intensity of electron emission from that area. Thus, for the
interpretation of the data it is important to know the physical processes leading to electron emission for a certain wavelength of the incident
light. The work function $\phi$ of Ag ranges between $4.2$ and $4.8$\,eV \cite{Michaelson77}, depending on crystal orientation. Upon
illumination of the sample with the UV lamp, electrons are emitted by regular, one-photon photoemission (1PPE), since $\hbar\omega_{UV}>\phi$.
On the other hand, under laser illumination at $\lambda=400~\mathrm{nm}~~\hbar\omega_{l}=3.1$\,eV, the photon energy is smaller than the work
function and photoemission requires a multiphoton process, where it can be expected that two-photon processes as lowest order dominate. In
two-photon photoemission (2PPE) the photoemission intensity is proportional to the forth power of the local electrical field which, especially
for LSP-resonant metal particles may significantly differ from the field of the incoming wave \cite{Merschdorf2004}. The presence of a Fermi
edge in the laterally resolved electron energy distribution spectra recorded from Cu and Ag nanoclusters \cite{Cinchetti04,Cinchetti04b}
demonstrate that even in this case 2PPE gives a substantial contribution to the recorded electron yield. As a first approximation it can be
assumed that in this case the electron emission yield scales with the square of near field photon density \cite{Shalaev96}, which is given by
the local electric field to the power 4. Due to the inelastic mean free path of the electrons, PEEM only probes the first few  nm (at our
energies about 5\,nm) from the surface, thus giving a fingerprint of the electrical field in this region. This quantity is crucial to understand
the aforementioned luminescence enhancement effects and is difficult to access by alternative near field imaging techniques. Details of the
interaction of the local electric field $\it E$ with the electrons certainly must take into account the vector character of $\it E$ as well as
the nature of the states from where electrons are emitted.

\begin{figure}
\begin{center}
\includegraphics[width=8cm,keepaspectratio=true]{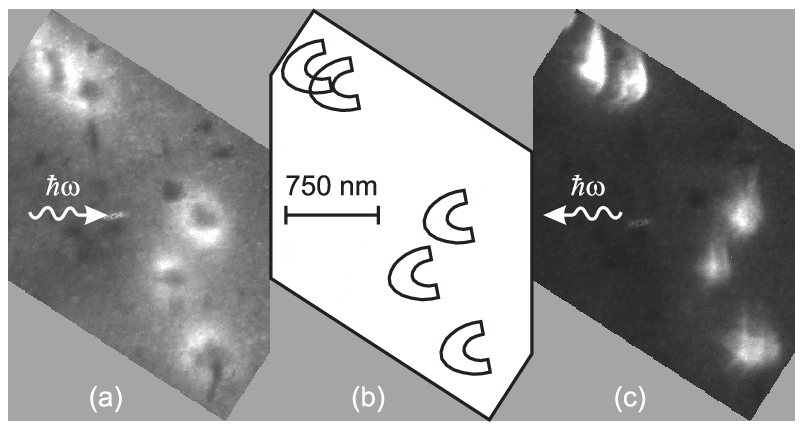}
\caption{High-resolution PEEM images of the same region of the sample.\\
(a) UV-PEEM image, $\hbar\omega_{UV}=5.0$\,eV (exposure time $\Delta t = 200~\mathrm{s}$); (b) Schematic drawing to indicate the position of the
silver crescents on the sample.
(c) Laser-PEEM image, $\hbar\omega_{l} =3.1$\,eV (exposure time $\Delta t = 5~\mathrm{s}$).\\
The arrows indicate the direction of illumination. Both images (a) and (c) have been digitally processed to enhance the contrast.}
\label{sample67_2}
\end{center}
\end{figure}
Figure \ref{sample67_2} (a) and (c) show  a UV-PEEM image and a laser-PEEM image of the same region of the sample, respectively. The
photoemission signal of five silver crescents is visible in both images. In particular, in (a) five ring-like structures can be identified. To
help the eye of the reader, we added Figure \ref{sample67_2} (b) to indicate the position and orientation of the crescents. In (c) the
photoemission yield is enhanced in different positions than in (a). Note that the dark spots visible in Figure \ref{sample67_2} (a) are due to
defects on the imaging unit of the photoemission electron microscope and must not be interpreted as part of the electron emission map.
\begin{figure}
\begin{center}
\includegraphics[width=7cm,keepaspectratio=true]{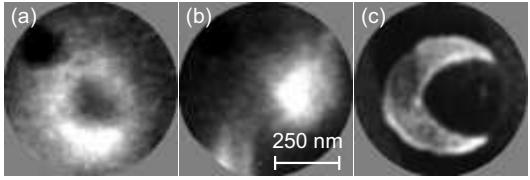}
\caption{ (a) Magnification of the third crescent from the top in Figure \ref{sample67_2} (a),
 $\hbar\omega_{UV} = 5.0$\,eV.\\
(b) Magnification of Figure \ref{sample67_2} (c), $\hbar\omega_{l} = 3.1$\,eV.\\
(c) Corresponding SEM image of a crescent with identical orientation and scale.} \label{NZ_scans}
\end{center}
\end{figure}
Figure \ref{NZ_scans} (a) and (b) show a magnification of the third crescent from the top in Figure \ref{sample67_2} (a) and (c) for the two
illumination modes. The dark spot in the upper left corner is one of the above mentioned defects. The images for UV and laser excitation reveal
a marked difference. In particular, comparison to the orientation of the crescents (c) suggests that upon UV illumination electron emission is
enhanced throughout the metal structure. Some highly localized features which are different for the individual objects are superimposed on this
average behavior. Illumination at $\lambda=400~\mathrm{nm}$ on the other hand leads to an enhanced emission in between the tips of the
structure. These observations can be explained by consideration of the dielectric response of silver \cite{Johnson72}. At the experimentally
used wavelengths silver has a dielectric function of $\epsilon\left(250~\mathrm{nm}\right) = -0.1377+3.5046\imath$ and
$\epsilon\left(400~\mathrm{nm}\right) =-4.4604+0.2147\imath$, The dominating imaginary part for $\epsilon\left(250~\mathrm{nm}\right)$ indicates
that this radiation corresponds to an energy above the onset of interband transitions. The dominating negative real part of
$\epsilon\left(400~\mathrm{nm}\right)$ is typical for all energies below the interband transitions: this behavior rules the entire frequency
range down to the static limit and may be termed "metallic" response.
\begin{figure}
\begin{center}
\includegraphics[width=6 cm,keepaspectratio=true]{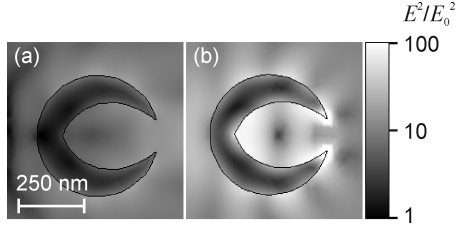}
\caption{Local magnitude of the electric field, calculated for a 2D geometry of silver in vacuum for light incident from the left with
wavelength $\lambda_{UV}=250~\mathrm{nm}$ (a) and $\lambda_{l}=400~\mathrm{nm}$ (b). The gray scale bar indicates the enhancement factor of the
squared field amplitude} \label{Calculation}
\end{center}
\end{figure}

Figure \ref{Calculation} shows calculations for a cross section through a silver rod in vacuum to illustrate qualitatively the optical near
field distribution for these two cases. At $\lambda=250~\mathrm{nm}$, there is an almost homogeneous field inside the silver while at
$\lambda=400~\mathrm{nm}$ enhanced and highly localized optical fields are observed, especially near to the tips.  This calculation must not be
regarded as a quantitative description of the electromagnetic response of the crescents since they are 50\,nm thick structures on an interface
between two media with highly different polarizabilities (vacuum and silicon) whereas the calculations are performed on infinitely extending
rods in vacuo. Still, the central conclusion of a qualitatively different response of the metal objects to optical fields above and below the
onset of interband transitions is justified and in agreement with theoretical studies on similar geometries
\cite{Kottmann01DoppelZylinder,Kottmann01PRB}.

As a general trend it can be stated that the particle plasmon wavelength given by $Re(\epsilon)=-2$ roughly divides a regime of metallic
behaviour at lower photon energies where large field enhancements and optical resonances are observed from a non-metallic regime, i.e.\ a
response without significant change of the field distribution of the exciting photon beam at higher energies. Our experimental observations can
be interpreted along these lines, homogeneous electron emission from the entire silver surface should appear as a 1:1 image of the geometrical
shape of the crescents for the case of UV illumination above the particle plasmon energy. This image is expected to be smeared out due to
convolution with the point spread function of the photoemission electron microscope, explaining the observation of ring-like structures in
Figure \ref{sample67_2} (a). The localized differences from ring to ring are partly due to imperfections in the crescents which may influence
the electron emission process in a way independent of the optical field strength (topographic contrast). It is noted, though, that for silver
particles the peculiar property of a blue-shifted plasmon frequency for very small particles is observed \cite{Liebsch93} which may point
towards another possible source for localized highly emissive spots at grains or cracks in the metal crescents in the UV. As a consequence of
these superimposed effects, the signature of the opening of the ring, which should be in principle visible, is highly obstructed. In the
laser-PEEM image (Figure \ref{NZ_scans} (b)), the enhanced 2PPE yield at the gap position points towards a locally enhanced electrical optical
field close to the tips of the structure in agreement with the behavior that can be expected for a photon energy in vicinity of  the particle
plasmon energy.

In summary, it has been shown experimentally for defined metal structures that local optical fields can be imaged directly by PEEM, a technique
which can reach a lateral resolution down to 20\,nm \cite{Ziethen98}. This provides an easy-to-use method for the quantitative investigation of
local field distributions ($\bf{\mathrm{E}}^2$ or $\bf{\mathrm{E}}^4$) at the surface. Note that our approach provides information that is
different from the light emission induced by electron beams which was reported already \cite{Yamamoto01}. In such experiments photons are
generated by electrons passing the optical antenna, such giving information on the 3D electrical field distribution around the antenna,
similarly to SNOM-experiments\cite{Frey02, Michaelis00, Hillenbrand2003, Krenn1999}. The PEEM method therefore promises to shed  more light on
the optical near field right at the metal surface, which up to now has been very difficult to access quantitatively.
%
%
It should be noted that description of the electromagnetic response in terms  of a macroscopic dielectric function $\epsilon \left(\omega
\right)$ can only be regarded as a first approximation near the surface of the material from where the photoelectrons are emitted. For the
metallic structure the microfields in this region are influenced by the oscillating surface charges, being located in the electron spill-out
region \cite{Liebsch93}. In the gap of the crescents the photoemission signal comes from the surface region of the oxide-covered silicon. Here,
the local field is modified by the properties of the vacuum-substrate interface. For both cases the near field cannot be treated properly using
an ansatz for a sharp interface between vacuum and material. It will be a challenging task for the future to develop a more refined model giving
a proper description of the field right at the surface. To reach this goal, we are also working on an optimized sample preparation with reduced
individual differences between the objects under study. Then, more experiments in combination with reliable theoretical predictions and
independent purely optical reference experiments are planned to achieve a complete understanding of the factors influencing the image formation.
\begin{acknowledgments}
We acknowledge financial support from BMBF projects 03N8702 and 03N6500.
\end{acknowledgments}
\bibliography{Bibliography}


%





\end{document}